\documentclass[12pt,a4paper]{article}%                	%
\usepackage{times}                                    	%
\usepackage{amsfonts,amsmath,amssymb}                 	%
\usepackage[a4paper]{geometry}                        	%
\usepackage{fancyhdr}                                 	%
\usepackage{color}                                    	%
\usepackage{multicol}
\usepackage[pdftex]{hyperref} % see note below
\usepackage{graphicx}%                                	%
\definecolor{c990000}{RGB}{153,0,0}
\hypersetup{                                          	%
a4paper,                                          	%
breaklinks                                        	%
}                                                     	%
{\ \rule{0.5em}{0.5em}}                               	%

\newcounter{ejmtFirstpage}                            	%
\usepackage[utf8]{inputenc}
\usepackage{xcolor}
\usepackage{breakurl}
\usepackage{csquotes}
\MakeOuterQuote{"}
\usepackage{float}
\usepackage{caption}
\begin{document}
%
% document title
%
\title{New tools in GeoGebra offering novel opportunities to teach loci and envelopes\thanks{GeoGebraBook
version of this paper is available at
\url{http://www.geogebratube.org/student/b128631}.}
}
%
% Single author.  Please supply at least your name,
% email address, and affiliation here.
%
\author{\begin{tabular}{cc}
\textit{Francisco Botana} & \textit{Zolt\'an Kov\'acs}\\
fbotana@uvigo.es & zoltan@geogebra.org\\
% The following line is too long, so I had to do some abbreviations instead:
%P\'apay Endre Vocational School & Department of Mathematics Education & Department of Analysis\\
Dept.~of Applied Mathematics I &The Priv.~Univ.~Coll.~of Educ.\\
Univ.~of Vigo, Campus A Xunqueira, Pontevedra&~of the Diocese of Linz\\
36005 & 4020\\
Spain & Austria\end{tabular}
}%

%
%%%%%%%%%%%%%%%%%%%%%%%%%%%%%%%%%%%%%%%%%%%%%%%%%%%%%%%%%%%
\date{}                                               	%
\maketitle                                            	%
%                                                     	%
%%%%%%%%%%%%%%%%%%%%%%%%%%%%%%%%%%%%%%%%%%%%%%%%%%%%%%%%%%%
%
% abstract
%
\begin{abstract}
GeoGebra is an open source mathematics education software tool being used in
thousands of schools worldwide. Since version 4.2 (December 2012) it
supports symbolic computation of locus equations as a result of joint 
effort of mathematicians and programmers helping the GeoGebra developer
team. The joint work, based on former researches, started in 2010 and
continued until present days, now enables fast locus and envelope 
computations even in a web browser in full HTML5 mode. Thus,
classroom demonstrations and deeper investigations of dynamic analytical
geometry are ready to use on tablets or smartphones as well.

In our paper we consider some typical secondary school topics where
investigating loci is a natural way of defining mathematical objects.
We discuss the technical possibilities in GeoGebra by using
the new commands {\bf LocusEquation} and {\bf Envelope}, showing through
different examples how these commands can enrich the learning of mathematics.
The covered school
topics include definition of a parabola and other conics in different
situations like synthetic definitions or points and curves associated with
a triangle. Despite the fact that in most secondary schools, no other than quadratic
curves are discussed, simple generalization of some exercises, and also every
day problems, will smoothly introduce higher order algebraic curves. Thus our paper
mentions the cubic curve ``strophoid'' as locus of the orthocenter of a
triangle when one of the vertices moves on a circle. Also quartic
``cardioid'' and sextic ``nephroid'' can be of every day interest when
investigating mathematics in, say, a coffee cup.

We also focus on GeoGebra specific tips and tricks when constructing a
geometric figure to be available for getting the locus equation. Among
others, simplification and synthetization (via the intercept theorem) are
mentioned.
\end{abstract}%
%
%%%%%%%%%%%%%%%%%%%%%%%%%%%%%%%%%%%%%%%%%%%%%%%%%%%%%%%%%%%
%                                                     	%
\thispagestyle{fancy}                                 	%
%                                                     	%
%%%%%%%%%%%%%%%%%%%%%%%%%%%%%%%%%%%%%%%%%%%%%%%%%%%%%%%%%%%
%
% Please use the following to indicate sections, subsections,
% etc.  Please also use \subsubsection{...}, \paragraph{...}
% and \subparagraph{...} as necessary.
%

\section{Introduction}
\subsection{Overview}
\href{http://www.geogebra.org}{GeoGebra} \textnormal{\cite{1}} is an open source mathematics
education software tool being used by millions of users worldwide. It is mainly
used to visualize mathematical relations in a dynamic way by supporting
reading correlations off not only visually but also numerically.

\begin{figure}[H]
\begin{center}
\includegraphics[width=120mm]{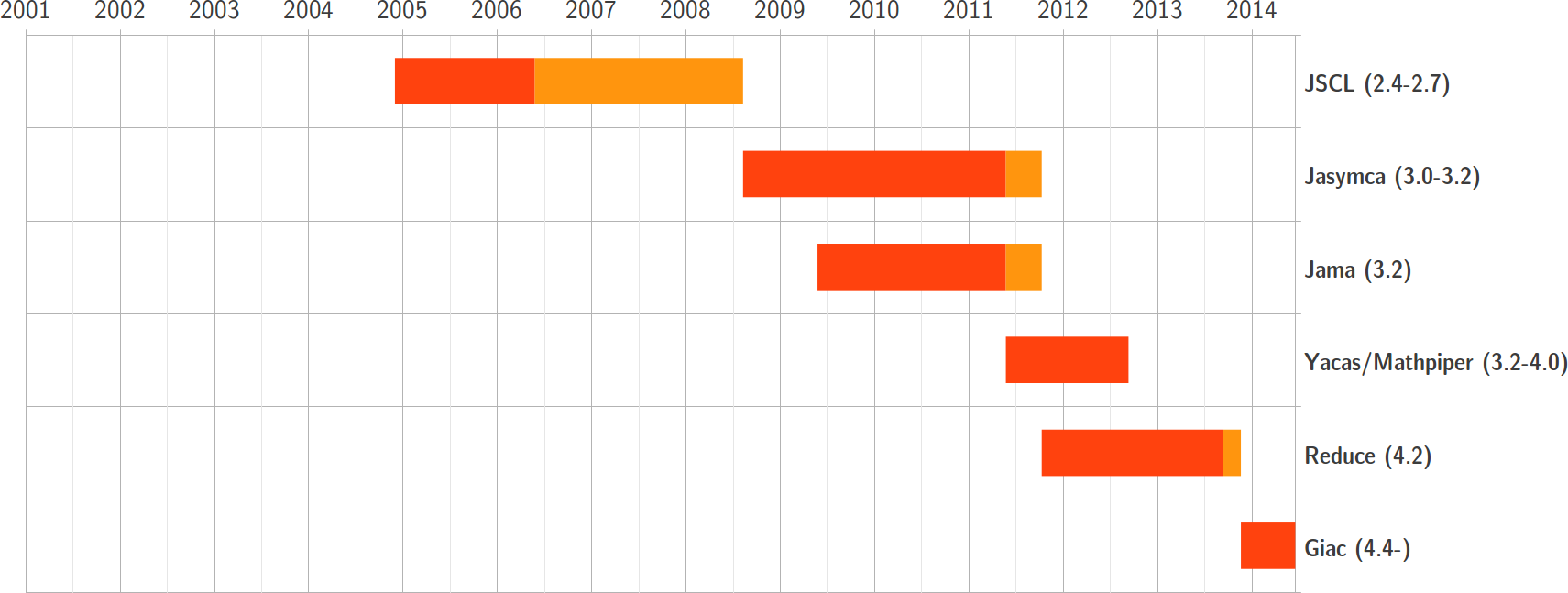}
\caption{Embedded CAS systems in various GeoGebra versions. Red period means continuous development phase in GeoGebra. 
Orange period shows stable phase with no longer development.}
\label{fig:cas}
\end{center}
\end{figure}

This approach has been continuously extended since 2004 by using an
embedded computer algebra system (CAS) in GeoGebra in version 2.4. In those
days CAS {\em JSCL} \textnormal{\cite{21}} was used which has been changed to {\em
\href{http://webuser.hs-furtwangen.de/~dersch/jasymca2/indexEN.html}{Jasymca} \textnormal{\cite{2}}}
(2008, GeoGebra 3.0), {\em
\href{http://math.nist.gov/javanumerics/jama/}{Jama} \textnormal{\cite{3}}} (2009, GeoGebra 3.2),
{\em Yacas/\href{http://www.mathpiper.org/}{Mathpiper} \textnormal{\cite{4}}} (2011, GeoGebra
3.2-4.0), {\em \href{http://www.reduce-algebra.com/}{Reduce} \textnormal{\cite{5}}} (2011,
GeoGebra 4.2) and {\em
\href{http://www-fourier.ujf-grenoble.fr/~parisse/giac.html}{Giac} \textnormal{\cite{6}}} (2013,
GeoGebra 4.4). (Fig.~\ref{fig:cas} illustrates the different CAS used by GeoGebra along the time.)

Interest in using CAS support in GeoGebra started a more specialized
interest in computing the algebraic equation of more general geometric
objects than lines and circles, namely locus equations. A team of
mathematicians (including the first author) located in Spain offered
scientific partnership and collaboration for the GeoGebra team (including
the second author) located in Austria. Their joint work was funded by
the {\em Google Summer of Code} program in 2010 by supporting the Spanish
university student {\em \href{http://www.serabe.com/tag/gsoc/}{Sergio
Arbeo}} to implement the computation of algebraic locus equations for
GeoGebra 4.0. Arbeo programmed the computations by using an extra CAS {\em
\href{http://krum.rz.uni-mannheim.de/jas/}{JAS} \textnormal{\cite{7}}} in GeoGebra, but his code
was later modified by the second author to use Reduce (and even later Giac)
instead.

As a result, locus equation computations are already present in GeoGebra
since version 4.2 and because of the numerous user feedback the newer
versions (including GeoGebra 5) contain some additional enhancements and
bug fixes as well. Also many users found the introduced {\bf LocusEquation}
command useful and easy to use in education as well. In Section 2 we
consider some possible classroom uses for the {\bf LocusEquation} command.

However Arbeo covered a wide set of classroom problems, meanwhile new
mathematical methods appeared to handle some problematic situations. The
\href{http://dx.doi.org/10.1016/j.cad.2014.06.008}{joint work} \textnormal{\cite{8}} of {\em
Montes}, {\em Recio,} {\em Abánades}, {\em Botana} and the {\em Singular}
CAS team yielded a powerful method to compute locus equations by
using the Gröbner cover (grobcov) package in Singular. In this way
GeoGebra has been extended to outsource computations to {\em
\href{https://code.google.com/p/singularws/}{SingularWS} \textnormal{\cite{9,sws-eaca}}}, an external web
service computing locus equations for GeoGebra (among other computations).
This method has been found extensible to compute not only locus
equations but envelopes as well. In Section 3 we show some of these
envelopes, pointing out the possibility of introducing them also in
secondary schools.

\section{Loci}
\subsection{A simple locus example}
A locus can be formally defined as a {\em set of points whose location
satisfies or is determined by one or more specified conditions}. Being more
specific, in GeoGebra a locus is the set of output points P' constructed by
given steps while the input point \textcolor[HTML]{7D7DFF}{P} is running on
a certain linear path. In other words, let point \textcolor[HTML]{7D7DFF}{P} be an
element of a linear path, and let point P' be the output point for the chosen
input \textcolor[HTML]{7D7DFF}{P} after some transformations of
\textcolor[HTML]{7D7DFF}{P} into P'.

In general the locus will be a curve, the set of output points P', since
the input points \textcolor[HTML]{7D7DFF}{P} also build up a curve. For example,
let the input curve be circle c and \textcolor[HTML]{7D7DFF}{P} a
perimeter point of c. Let the center of the circle be
\textcolor[HTML]{0000FF}{C}. Now let us construct point P' such that P' is
the midpoint of \textcolor[HTML]{7D7DFF}{P}\textcolor[HTML]{0000FF}{C}.
Clearly, the locus curve here is also a circle described by center
\textcolor[HTML]{0000FF}{C} and half of the radius of c (Fig.~\ref{fig:2}).

This example can be entered into GeoGebra either by using the graphical
user interface with the mouse  (as it is traditionally done in dynamic geometry environments),
or by the keyboard in the Algebra Input
(here we put point \textcolor[HTML]{0000FF}{C} to $(2,3)$ and use radius
4):

\begin{itemize}
\item {\bf \textcolor[HTML]{0000FF}{C}=(2,3)}
\item {\bf c=Circle[\textcolor[HTML]{0000FF}{C},4]}
\item {\bf P=Point[c]}
\item {\bf P'=Midpoint[\textcolor[HTML]{7D7DFF}{P},\textcolor[HTML]{0000FF}{C}]}
\item {\bf Locus[P',\textcolor[HTML]{7D7DFF}{P}]}
\end{itemize}

It should be noted that saying that the locus curve is a circle, is, or at least could be, a risky
statement. There is no immediate reason, besides the visual one in Fig.~\ref{fig:2}, to conclude that the locus
is a circle. Since it is well known that visualization is prone to errors, there is no doubt about the
interest on having a certified method to get the equation of the locus curve. As said above,
GeoGebra now incorporates a protocol for it. Replacing the command {\bf Locus[P',\textcolor[HTML]{7D7DFF}{P}]}
by {\bf LocusEquation[P',\textcolor[HTML]{7D7DFF}{P}]} we can also check the result
algebraically: an implicit curve is displayed with the equation
$x^2-6x+y^2-4y=-9$. 
Although the curve is not identified as a circle, knowing its equation is a capital step for classifying
it. Furthermore, it is also possible to see how the equation changes
dynamically when point \textcolor[HTML]{0000FF}{C} is dragged.

How can this equation be computed mathematically? Let us define coordinates
$x_C$, $y_C$, $x_{P}$, $y_P$, $x_{P'}$ and $y_{P'}$ for the points defined
above. Now the following equations are valid:

\begin{enumerate}
\item $x_C=2$
\item $y_C=3$\item $(x_P-x_C)^2+(y_P-y_C)^2=4^2$\item
$x_{P'}=\frac{x_P+x_C}{2}$\item $y_{P'}=\frac{y_P+y_C}{2}$
\end{enumerate}

What we need is to reduce this equation system to a single equation
containing only coordinates of point P'. In algebra this computation is
called {\em elimination}, i.e.~we should eliminate all variables except $x_{P'}$ and
$y_{P'}.$

In GeoGebra this computation is achieved by the Giac CAS in the background,
but it can also be computed directly by using a GeoGebra command (which
calls the appropriate Giac statement):

\begin{figure}[H]
\begin{center}
\fbox{\includegraphics[width=120mm,height=63.6mm]{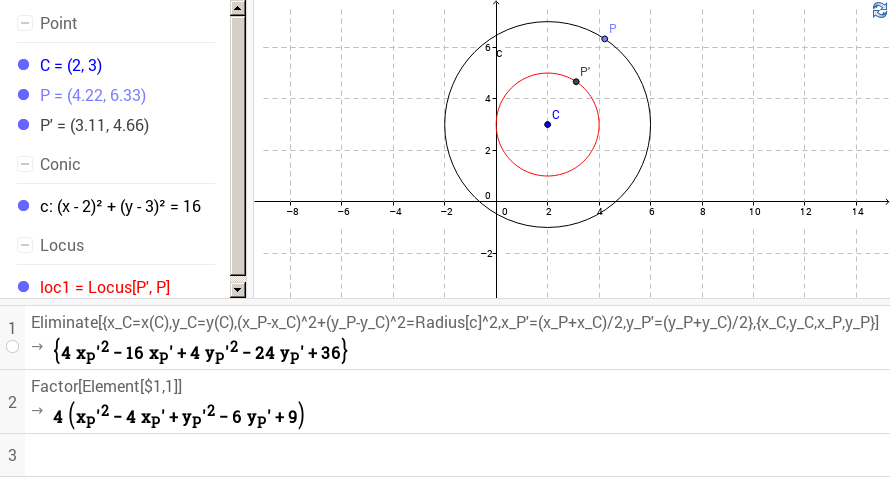}}
\caption{A simple GeoGebra locus.}
\label{fig:2}
\end{center}
\end{figure}

Here the user can add command {\bf
LocusEquation[\textcolor[HTML]{FF0000}{loc1}]} to make GeoGebra compute the
equation automatically. By clicking on the marble to the left of the input
line GeoGebra will also display the geometric form of the equation, i.e.~another
circle will be drawn (the same as \textcolor[HTML]{FF0000}{loc1}).

\subsection{Parabolas as loci}

Apart from straight lines and circles, the parabola is probably the most used object when illustrating
the concept of locus. In this section we recall parabolas following their traditional definition and
through the path described by a triangle center when a vertex is dragged. Furthermore, we show, in
the second example, how simple situations can lead to generalizations.

\subsubsection*{Definition of a parabola}

The common parabola definition, {\em the set of points
equidistant from a single point (the focus) and a line (the directrix)},
is, however, not easy to handle for some students. There is at
least one abstract step in-between, namely that to find the distance from a
line we may need a perpendicular being drawn.

Thus, when focus \textcolor[HTML]{0000FF}{F} and directrix d are given,
constructing one point P of parabola {\color{red}p} is as follows:

\begin{enumerate}
\item Choose an arbitrary point \textcolor[HTML]{7D7DFF}{D} of
\textcolor[HTML]{0000FF}{d}.\item Construct a perpendicular line to
\textcolor[HTML]{0000FF}{d} on \textcolor[HTML]{7D7DFF}{D}.\item Construct
the bisector b of \textcolor[HTML]{7D7DFF}{D} and
\textcolor[HTML]{0000FF}{F}.\item Let the intersection point of the
perpendicular line and b be P.\item Now P is a point of parabola
\textcolor[HTML]{FF0000}{p} since
P\textcolor[HTML]{7D7DFF}{D}=P\textcolor[HTML]{0000FF}{F} (because bisector
b is actually the axis for the mirroring of point
\textcolor[HTML]{7D7DFF}{D} to \textcolor[HTML]{0000FF}{F}).
\end{enumerate}

In fact drawing bisector b is also a hidden step since we implicitly used
some basic properties of the reflection.

This kind of definition of the parabola is usual in many secondary schools,
however the equivalence of this definition and the analytical one (that is,
the usual formula for a parabola is $y=ax^2+bx+c$ for some constants $a$, $b$
and $c$) is not obvious.

\begin{figure}[H]
\begin{center}
\includegraphics[width=100mm]{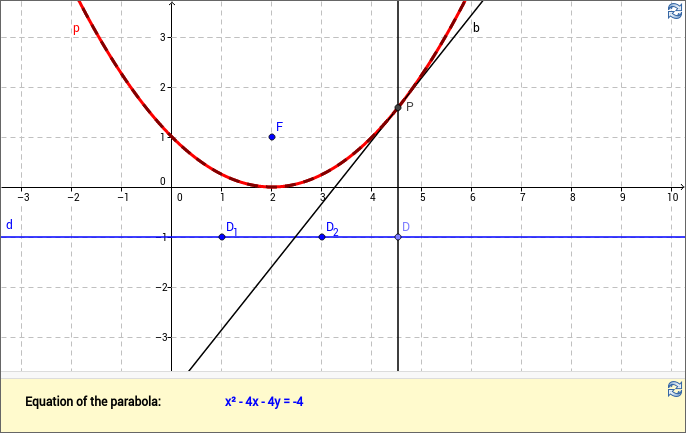}
\caption{A parabola obtained as locus and its equation.}
\label{fig:3}
\end{center}
\end{figure}

In the Figure \ref{fig:3} we can see a dark red dashed parabola which is the real locus
drawn by GeoGebra numerically. The lighter red curve has been computed
symbolically by the {\bf LocusEquation} command. While in this case these
two curves are exactly the same, we will see some examples below where it
is not the case. The reason comes from the algorithm we use to compute the
equation. (In many cases the symbolical result can be improved by using
extended algorithms.)

\subsubsection*{Locus of the orthocenter}

\begin{figure}[H]
\begin{center}
\includegraphics[width=107mm]{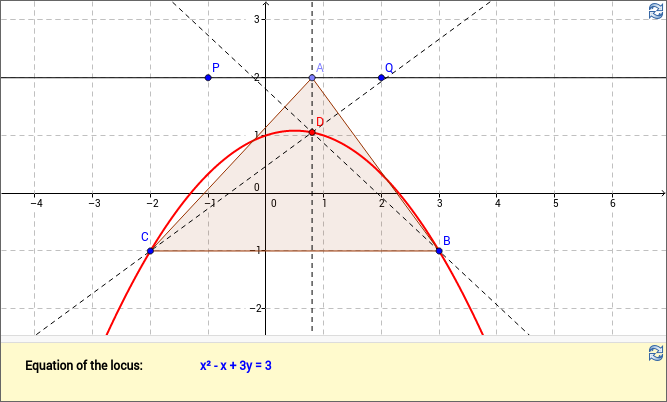}
\caption{Locus of the orthocenters of a triangle.}
\label{fig:4}
\end{center}
\end{figure}

A parabola can also be obtained by getting the orthocenter points of a
triangle if two vertices are fixed and the third one moves on a line which
is parallel to the opposite side (see Fig.~\ref{fig:4}, where vertex
\textcolor[HTML]{7D7DFF}{A} moves along line \textcolor[HTML]{0000FF}{PQ},
parallel to triangle side {\color{blue}BC}).

In this figure point \textcolor[HTML]{7D7DFF}{A} is constrained by line
\textcolor[HTML]{0000FF}{PQ}. Students can drag point
\textcolor[HTML]{7D7DFF}{A} on line \textcolor[HTML]{0000FF}{PQ} and see
how the orthocenter {\color{red}D} is changing meanwhile. It is not obvious
to prove that the locus here is a parabola, but the students are able at
least to get experience by changing points {\color{blue}B} and
{\color{blue}C} by preserving parallelism of lines
\textcolor[HTML]{0000FF}{PQ} and \textcolor[HTML]{0000FF}{BC}.

On the other hand, the locus equation will not be essentially different on
other positions of {\color{blue}P}, {\color{blue}Q}, {\color{blue}B} and
{\color{blue}C}: it will be quadratic in most sets of positions. Some of
these positions seem easy to investigate, for example when
\textcolor[HTML]{0000FF}{PQ} is perpendicular to
\textcolor[HTML]{0000FF}{BC} (here the result will be a linear equation).
Others, for example by putting {\color{blue}P} to $(1,1)$ and not changing
anything else in the set, the locus result is a hyperbola, namely
$x^2+xy+y=5$. First, this formula is hard to analyze in secondary school
since it is not in explicit form like a function $y=f(x)$. Second, this
formula is still a quadratic implicit equation and thus it can open
horizons of generalization to cover all kind of conics.

In case {\color{blue}P}=$(1,1)$, {\color{blue}Q}=$(2,0)$,
{\color{blue}B}=$(3,-1)$, {\color{blue}C}=$(3,1)$ the computed locus equation is
$-xy-x+y^2+3y=-2$, i.e. $-xy-x+y^2+3y+2=0$ whose left hand side is the
product of $(y+1)$ and $(y-x+2)$, two lines, namely $y=-1$ and $y=x-2$
written in the usual explicit form. In this constellation altitude of side
\textcolor[HTML]{7D7DFF}{A}{\color{blue}B} always lies on line
\textcolor[HTML]{0000FF}{CQ} since it is perpendicular to
\textcolor[HTML]{0000FF}{PQ}. Thus point {\color{red}D} will also lie on
line \textcolor[HTML]{0000FF}{CQ}, so it seems sensible that the locus
equation is \textcolor[HTML]{0000FF}{CQ} in this case. Unfortunately,
GeoGebra's {\bf LocusEquation} shows an extra line here, not only
\textcolor[HTML]{0000FF}{CQ} (which has the explicit equation $y=x-2$) but
also another one. This example shows that the real locus may be a subset of
the result of the {\bf LocusEquation} command.

\begin{figure}[H]
\begin{center}
\includegraphics[trim=0 0 10cm 0,clip,width=12cm]{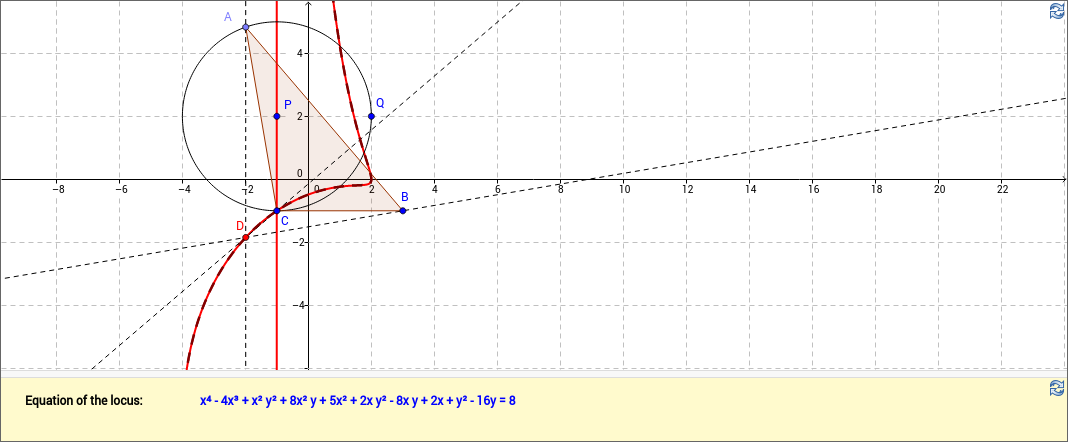}
\caption{A strophoid obtained as a locus (with an extra linear factor).}
\label{fig:5}
\end{center}
\end{figure}

One further step forward is to constrain point \textcolor[HTML]{7D7DFF}{A}
on a circle, not a line. In this way we can obtain non-quadratic locus
equations like the strophoid formula which is a cubic one (see Fig.~\ref{fig:5}
where \textcolor[HTML]{7D7DFF}{A}
is bound to a circle centered at \textcolor[HTML]{0000FF}{P} and passing through
\textcolor[HTML]{0000FF}{Q}, being both
\textcolor[HTML]{0000FF}{P} and \textcolor[HTML]{0000FF}{Q}
user defined points).

In this figure a quartic equation is shown, but the real locus is a cubic
curve. That is, an extra component is shown (here line $x=-1$) as in many
other cases when dragging points \textcolor[HTML]{0000FF}{P},
\textcolor[HTML]{0000FF}{Q}, \textcolor[HTML]{0000FF}{B} or
\textcolor[HTML]{0000FF}{C}. On the other hand, by moving point
\textcolor[HTML]{0000FF}{P} down, for example to $(-1,1)$ or $(-1,0)$, the
locus is the same as {\bf LocusEquation} computes: in these cases the locus
is a real quartic curve.

A beautiful side case is when {\color{blue}B}={\color{blue}P}
and {\color{blue}C}={\color{blue}Q}. In this case the real locus is a
\href{http://mathworld.wolfram.com/RightStrophoid.html}{right strophoid} \textnormal{\cite{10}}
curve and an extra line component is drawn on points {\color{blue}C} and
{\color{blue}Q}.
Concerning the extra components when
finding loci, we refer to \cite{8} where a complete theoretical solution is given.

\subsection{Technical details}
\subsubsection*{Computational background}

We think that most readers of this papers are neither technicians, nor
mathematicians, but teachers. Nevertheless, it is good to know some
computational details how the {\bf LocusEquation} command works.
There should be a compromise between using educational software as a blackbox and being
an expert developer.

Computing a locus equation can be time consuming even for fast computers.
Basically, a set of equations has to be created in the background: the more
objects we have in our construction, the more variables and equations we
need. After setting up algebraic equations, they have to be solved
symbolically in an efficient way. For this task we use {\em Gröbner bases} \cite{18}.

GeoGebra uses the Giac computer algebra system to compute Gröbner bases as
efficiently as possible, but the general method is still double
exponentional in the number of variables. On the other hand, Giac runs in a
web browser in today's computers, and this slows down computations by
almost one magnitude. (This means that computing locus equation in the
desktop version of GeoGebra is still much faster than observing the
construction in a web browser. For the future, however, there are plans to
speed up JavaScript computations by substituting them with native
instructions.)

Giac is a powerful CAS, but it can slow down if extreme input must be
processed. That is why it is desired to solve equations only having integer
coefficients. To achieve this, it is suggested using so-called dynamic
coordinates in GeoGebra: to create free points \textcolor[HTML]{0000FF}{A'},
{\color{blue}B'}, {\color{blue}C'}, $\ldots$ first and then define point {\bf
A=DynamicCoordinates[{\color{blue}A'},round(x({\color{blue}A'})),round(y({\color{blue}A'}))]},
then use similar definitions for points B and C and so on.

A final remark on computational issues deals with the type of numbers involved in calculi.
Using "easy" coordinates will speed up computations. For example,
putting A to the origin, B on the $x$-axis and using small integers instead
of larger numbers may decrease computation time significantly.

\subsubsection*{Supported construction steps}

Since the Gröbner basis computation assumes algebraic (polynomial)
equations, there are restrictions for the available construction steps for
the {\bf LocusEquation} command. First of all, only Euclidean construction
steps are supported. Even if a step could be converted into a Euclidean
construction, some non-trivial way of wording are not supported, for
example, when the user defines a parabola by entering its explicit formula,
then it cannot be discovered by the {\bf LocusEquation} command at the
moment. Instead, the parabola must be constructed by using the appropriate
GeoGebra tool.

Most Euclidean two dimensional construction steps are already supported. 
Javadoc at
\href{http://dev.geogebra.org/trac/wiki}{the GeoGebra Developer Wiki} \textnormal{\cite{11}}
provides a full list of them.

If a geometry problem is described fully or partially by formulas, it may
be difficult to translate it to a purely Euclidean construction. Below we
will see an example how this can be achieved.

\subsubsection*{Agnesi's witch}

Here we provide two examples to implement
\href{http://en.wikipedia.org/wiki/Witch_of_Agnesi}{Agnesi's witch} \textnormal{\cite{12}} in
GeoGebra. The first approach is a general way which will result in
slow computation and some extra components. The second approach is
much faster and results only in one extra component.

For the first approach we simply consider the formula $y=\frac{1}{x^2+1}$.
Here we need to define the unit ($1$) and compute the square of $x$ based on
this unit, then add these two lengths. Then we need to compute the
reciprocal of the result, and translate the final length $y$ to the correct
position of the coordinate system.

In Figure \ref{fig:6} we can see a numerical locus in red and a symbolical locus
in blue. In fact Agnesi's curve is just a part of these curves since the
conversion of its formula will introduce extra components. Let us follow
the steps we made in this figure:

\begin{figure}[H]
\begin{center}
\includegraphics[width=120mm]{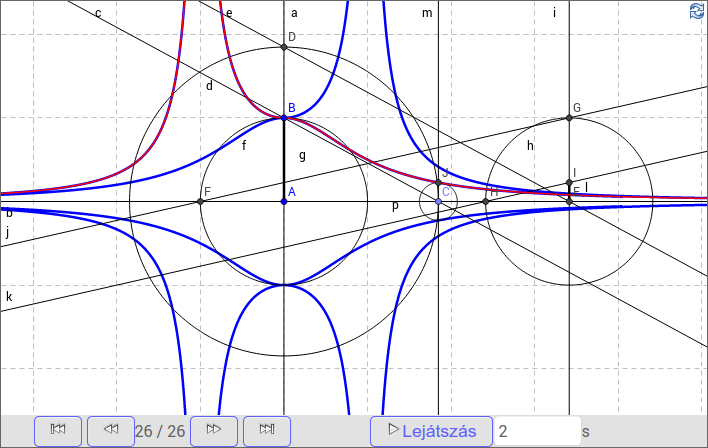}
\caption{A locus where the witch of Agnesi is contained.}
\label{fig:6}
\end{center}
\end{figure}

\begin{enumerate}
\item Point \textcolor[HTML]{0000FF}{A} is created (origin).\item Point
\textcolor[HTML]{0000FF}{B} is created, $(0,1)$.\item Line a is the
$y$-axis.\item Line b is the $x$-axis.\item Point \textcolor[HTML]{7D7DFF}{C}
lies on the $x$-axis, it will be the projection of a point of the curve to
the $x$-axis, i.e. its abscissa will be $x$.\item We will use the
\href{http://en.wikipedia.org/wiki/Intercept_theorem}{intercept theorem} \textnormal{\cite{13}} to
construct $x^2$. So we create a triangle with sides having length $1$
(\textcolor[HTML]{0000FF}{AB}) and $x$
(\textcolor[HTML]{0000FF}{A}\textcolor[HTML]{7D7DFF}{C}). This triangle
is right, but this property is not necessary. The third side of the
triangle is line c.\item For the intercept theorem we prepare length $x$
also on line {\color{blue}AB} by drawing circle d.\item Point D is
intersection of line a and circle d. (In fact there are two intersection
points here, but we use the "upper" one.) Now {\color{blue}A}D=$x$.\item
Line e is parallel to c and lies on D.\item Point E is intersection of lines
b and e. By using the intercept theorem obviously
{\color{blue}A}E=$x^2$.\item We are preparing addition, thus we draw
another circle f around the origin having unit radius.\item Point F is
intersection of line b and circle f. (In fact there are two intersection
points here, but we use the "left" one.) Now EF=$x^2+1$.\item We would like
to copy the unit length, so we create segment g as the unit (i.e.,
AB).\item Circle h is around point E with unit radius.\item Line i is
perpendicular to line b and lies on point E.\item Point G is intersection
of line i and circle h. (In fact there are two intersection points here,
but we use the "upper" one.) Now GE is a copy of the unit. This is a
preparation for applying another intercept theorem.\item We will use
triangle EFG for the intercept theorem, thus we draw line FG as line
j.\item Point H is again an intersection of circle h and line b. (The
"left" one.) Now we copied the unit as EH, too.\item Drawing line k as a
parallel one with line j through H.\item Intersection of lines i and k is
point I.\item Now applying the intercept theorem for length IE we obtain
IE=$\frac{1}{x^2+1}$. This will be $y$.\item Now we will copy this length to
point \textcolor[HTML]{7D7DFF}{C} orthogonally. Thus we draw a
perpendicular with line b through \textcolor[HTML]{7D7DFF}{C}. This is
line m.\item We copy length IE to point \textcolor[HTML]{7D7DFF}{C}
upwards, thus we draw circle p around \textcolor[HTML]{7D7DFF}{C} with
radius IE.\item "Upper" intersection point of m and p is point J.\item
Locus of point J while \textcolor[HTML]{7D7DFF}{C} is moving on the $x$-axis
is what we search for. In fact, only those points J are proper which have
positive abscissa.\item Finally, {\bf LocusEquation} shows a $12$ degree
polynomial, $x^8y^4-2x^4y^4-2x^4y^2+y^4-2y^2+1$, which is a product of the
cubic Agnesi curve and its reflection to the $x$-axis, and two other cubics
(reflections of each other), namely $x^2y-y-1$, $x^2y-y+1$, $x^2y+y-1$ and
$x^2y+y+1$.
\end{enumerate}

After finishing this construction it is clear that we almost surely obtain
extra components since it is impossible to exclude the "right" intersection
point in steps 12 and 18, for example. The construction process described
above illustrates the weakness of the Euclidean (i.e.~in fact algebraic)
method.

\begin{figure}[H]
\begin{center}
\includegraphics[width=80mm]{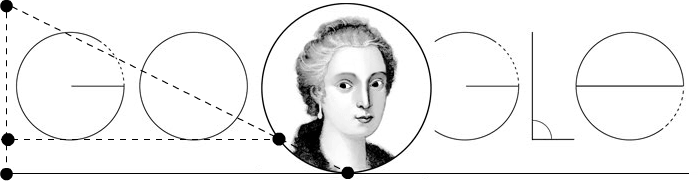}
\caption{Agnesi's witch as a Google doodle.}
\label{fig:7}
\end{center}
\end{figure}

Finally we refer to a simpler definition of Agnesi's curve, also used
by a Google "doodle" on the 296th anniversary of Maria Agnesi's birth on 16
May 2014 (Fig.~\ref{fig:7}).
Given the center {\color{blue}A} (here the origin) and circumpoint {\color{blue}B}
(here the point $(0,-2)$) of a circle, let {\color{blue}B}\textcolor[HTML]{7D7DFF}{C} the diameter of the circle. Let
\textcolor[HTML]{7D7DFF}{b}
the tangent line of the circle at \textcolor[HTML]{7D7DFF}{C}, and let \textcolor[HTML]{7D7DFF}{D} be an arbitrary point on the circle.
Now E is the
intersection point of {\color{blue}B}\textcolor[HTML]{7D7DFF}{C} and
\textcolor[HTML]{7D7DFF}{b}. At last {\color{red}F} will be the projection of E on a line parallel to the $x$-axis,
going through \textcolor[HTML]{7D7DFF}{D}.

\begin{figure}[H]
\begin{center}
\includegraphics[width=80mm]{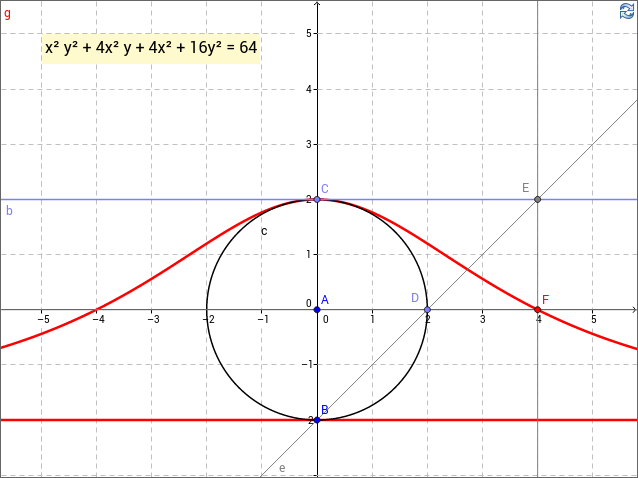}
\caption{The witch of Agnesi as a simpler locus.}
\label{fig:8}
\end{center}
\end{figure}

This "easy" definition allows GeoGebra to show Agnesi's witch much faster
than above and make it computationally possible to drag the input points
even in a web browser (Fig.~\ref{fig:8}). Of course, in many cases such a simplification is an
intellectual challenge by searching for algebraic or geometric
simplifications to result in less variables in the Gröbner basis
computations.

As an exercise, we leave to the reader to prove that the trace of point
\textcolor[HTML]{FF0000}{F} is $y=\frac{1}{x^2+1}$ if
\textcolor[HTML]{0000FF}{B} is in the origin,
\textcolor[HTML]{0000FF}{A}=$(0,1/2)$ and \textcolor[HTML]{7D7DFF}{C}=$(0,1)$.
Also as an exercise in GeoGebra to improve this figure: use {\bf
DynamicCoordinates} instead of point capturing to grid points (which yields
non-continuous motion for point \textcolor[HTML]{7D7DFF}{D}). Another
improvement can be to put the equation text into a fix position, preferably
in the second Graphics View.

\section{Envelopes}

We have studied loci of points in the preceding section. Replacing points by lines, it arouses the
concept of envelope. Through this section we consider these objects, that is, the locus of a family of
plane lines depending on some other object. To this end, we describe how a parabola can be
obtained by studying a family of tangent lines, we recall a kind of optical art relating it with tangent
lines, and, finally, we illustrate how mathematics can be learned in a cup of coffee.

\subsection{Motivation}
\begin{figure}[H]
\begin{center}
\includegraphics[width=56.2mm]{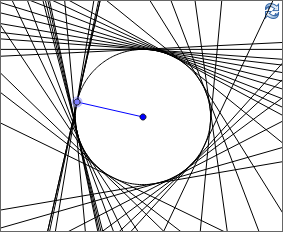}
\caption{A circle suggested by tracing tangent lines to its points.}
\label{fig:9}
\end{center}
\end{figure}

A simple definition of a circle is {\em a plane
curve everywhere equidistant from a given fixed point, the center}. Now by
constructing the tangent line in a point of the circle we find that the
tangent is perpendicular to the radius. When considering the trace of the
tangent lines as shown in Fig.~\ref{fig:9}, we find that the union $U$ of the tangents of a circle is the
whole plane except the disc inside the circle.

Now let us consider the same figure from another point of view. Having the
set $U$ we may be interested of a curve such that its tangents are the lines
of $U$. Such a curve can be the given circle, but that it is the only
possible curve (that is, the question has a unique answer) is not
straightforward.

\begin{figure}[H]
\begin{center}
\includegraphics[width=51.6mm]{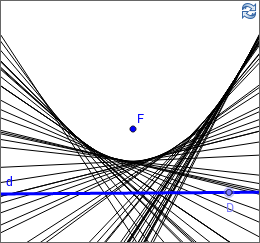}
\caption{Tracing lines to suggest a parabola.}
\label{fig:10}
\end{center}
\end{figure}

On the other hand, each element of $U$ is equidistant from the center of the
circle because distance is defined by measuring orthogonal projection. This
idea leads us to define a parabola by considering the set $U'$ of lines being
equidistant from a given point \textcolor[HTML]{0000FF}{F} and a given line
\textcolor[HTML]{0000FF}{d}. To measure the distance from line
\textcolor[HTML]{0000FF}{d} we consider each point
\textcolor[HTML]{7D7DFF}{D} of line \textcolor[HTML]{0000FF}{d} and take
the perpendicular bisector of points \textcolor[HTML]{0000FF}{F} and
\textcolor[HTML]{7D7DFF}{D} as shown in Fig.~\ref{fig:10}.

We have already mentioned in the previous section that usual definition of
a parabola p' is that it is the locus of points P' which are equidistant
from focus \textcolor[HTML]{0000FF}{F'} and directrix d'. Now let
\textcolor[HTML]{7D7DFF}{D' } be an arbitrary point of d', line b the
bisector of segment \textcolor[HTML]{0000FF}{F'}\textcolor[HTML]{7D7DFF}{D'
} and P' the intersection of the perpendicular to d' in
\textcolor[HTML]{7D7DFF}{D' } and b. A well-known property of b is that it is
the tangent of parabola p' in point P'.

\begin{figure}[H]
\begin{center}
\includegraphics[width=91.8mm]{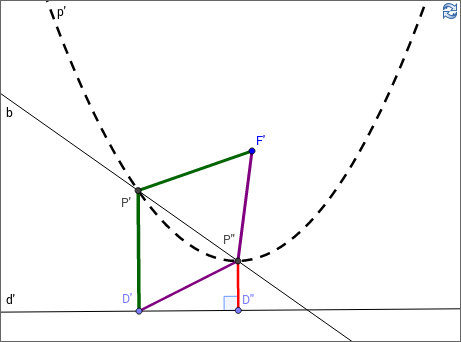}
\caption{Line b is not secant to parabola, but tangent.}
\label{fig:11}
\end{center}
\end{figure}

Let us assume that b is not a tangent of parabola p' in point P'. Then
there is another point P'' on b, also element of the parabola, that is, b is
a secant line of p'. Assumably this point P'' was created by foot point
\textcolor[HTML]{7D7DFF}{D''} (element of d'), for which
\textcolor[HTML]{7D7DFF}{D''}P''=P''{\color{blue}F'}. Since b is the
bisector of \textcolor[HTML]{7D7DFF}{D'}\textcolor[HTML]{0000FF}{F'}, also
\textcolor[HTML]{7D7DFF}{D'}P''=P''\textcolor[HTML]{0000FF}{F'} holds. But
this means that
\textcolor[HTML]{7D7DFF}{D'}P''=\textcolor[HTML]{7D7DFF}{D''}P'', that is,
in triangle \textcolor[HTML]{7D7DFF}{D'}\textcolor[HTML]{7D7DFF}{D''}P''
(which is a right triangle) hypothenuse \textcolor[HTML]{7D7DFF}{D'}P'' and
cathetus \textcolor[HTML]{7D7DFF}{D''}P'' have the same length, which is
impossible. This contradiction ensures that b is a tangent, not a secant,
as shown in Fig.~\ref{fig:11}.

Thus we proved that these two different definitions of a parabola (i.e.~the
classical one by using locus, and this second one which uses the concept of
trace of the bisector) are equivalent, that is, they define the same
parabola.

\subsection{Envelopes and string art}
{\em String art}, or pin and thread art, according to
\href{http://en.wikipedia.org/wiki/String_art}{Wikipedia} \textnormal{\cite{14}}, is characterized
by an arrangement of colored thread strung between points to form abstract
geometric patterns or representational designs such as a ship's sails,
sometimes with other artist material comprising the remainder of the work.

\begin{figure}[H]
\begin{center}
\includegraphics[width=60mm]{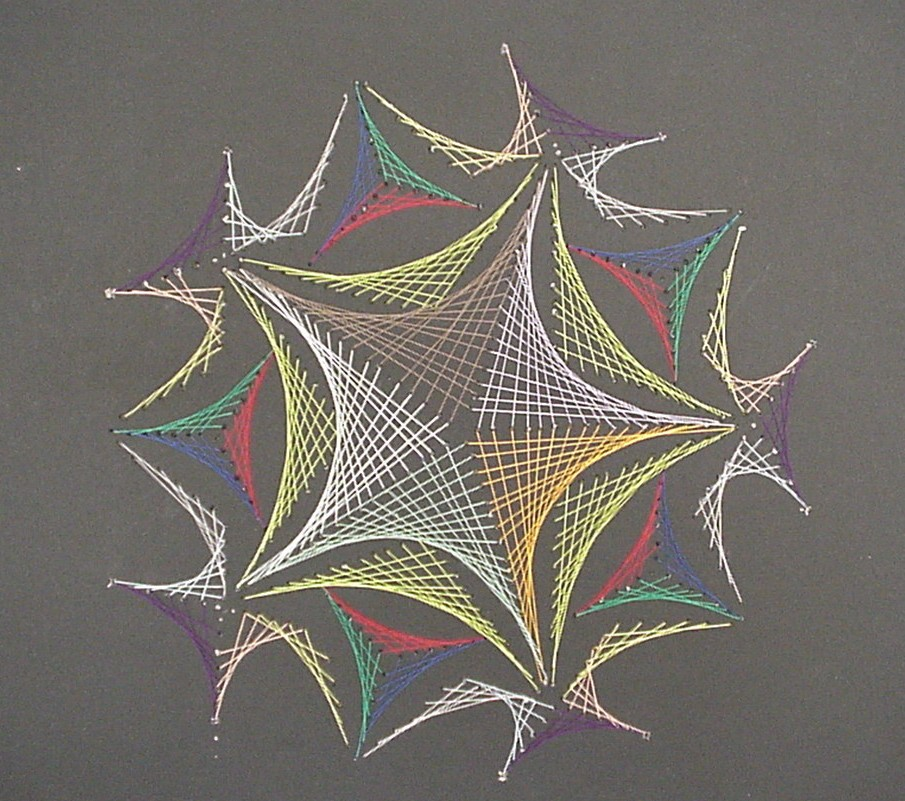}
\caption{An example of string art.}
\label{fig:12}
\end{center}
\end{figure}

In Figure \ref{fig:12} strings play the same role as tangents in the previous
examples. As Markus Hohenwarter, inventor of GeoGebra refers in paper
\textnormal{\cite{20}} and in the 8th chapter of
\textnormal{\cite{15}}, the segments we can see are tangents to a quadratic Bézier curve.

\begin{figure}[H]
\begin{center}
\includegraphics[width=60mm]{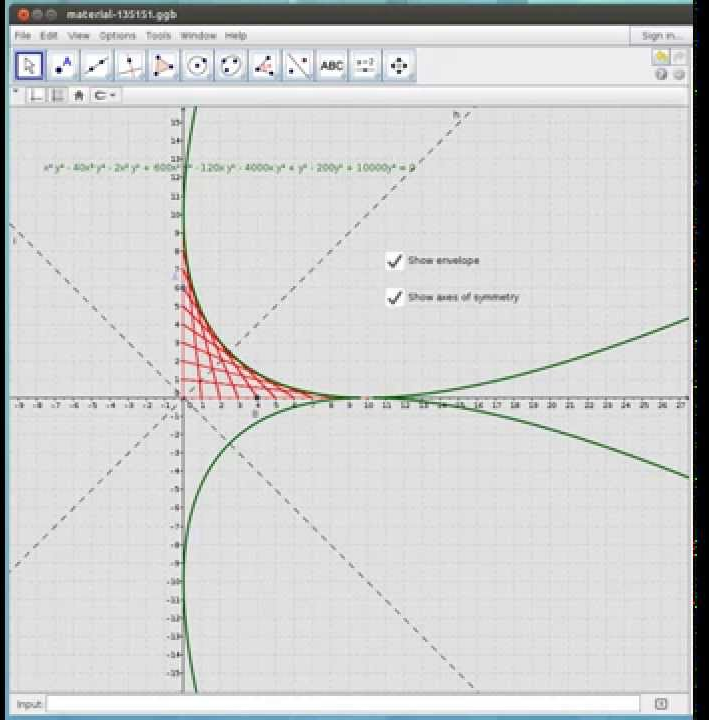}
\caption{Checking the grade of a contour curve. YouTube video at \url{https://www.youtube.com/watch?v=V-Cq2VMsiZw}.}
\label{fig:13}
\end{center}
\end{figure}

In the following \href{http://geogebratube.org/student/m135151}{applet} \textnormal{\cite{16}} one
can do a similar experiment by using GeoGebra, and eventually use its {\bf
Envelope} command to check whether the resulted contour curve is quadratic (Fig.~\ref{fig:13}
shows a capture of the mentioned applet where the grade of the equation is 2).

By no mean the activity to draw the segments \textcolor[HTML]{7D7DFF}{A}B
while \textcolor[HTML]{7D7DFF}{A} is dragged on grid points between $(0,0)$
and (0,10)---and meanwhile B is moved between $(10,0)$ and $(0,0)$---is an easy
task for many types of school pupils. Also the result as getting the
contour of the segments can be expected to be a straightforward way of the
next step of understanding. However, obtaining the envelope equation is at
a different step of difficulty level.

First of all, the obtained equation is of 5th grade, containing not only
the curve itself, but its reflection to the $x$-axis, and also the $x$-axis
itself. This equation is
$x^4y-40x^3y-2x^2y^3+600x^2y-120xy^3-4000xy+y^5-200y^3+10000y=0$ which is
an implicit equation, but it can be factorized into three factors.
Unfortunately, factorization is not discussed at secondary level, so we
need to find another approach to go into the very details.

Fortunately, all these problems can be managed in secondary school by
changing the construction in some sense. On one hand, we rotate the
axes by 45 degrees to obtain an explicit equation: in this case one of the
parabolas can be written in form $y=ax^2+bx+c$. On the other hand, we
use magnification of 10, so that we obtain parabola
$y=\frac{x^2}{2}+\frac{1}{2}$ which can be described with directrix $y=0$
and focus $(0,1)$.

\begin{figure}[H]
\begin{center}
\includegraphics[width=109mm]{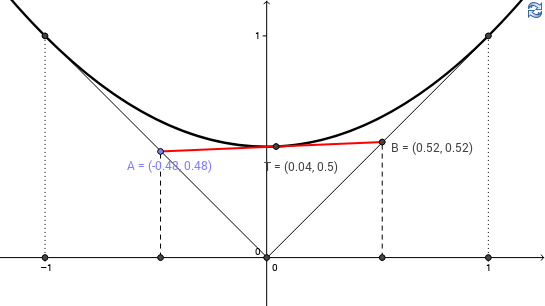}
\caption{Another construction of a parabola.}
\label{fig:14}
\end{center}
\end{figure}

Now we prove that segment \textcolor[HTML]{7D7DFF}{A}B is always a tangent
of the parabola described in Figure \ref{fig:14}. We use only such methods which can be
discussed in a secondary school as well. We would like to compute the
equation of line \textcolor[HTML]{7D7DFF}{A}B to find the intersection
point T of \textcolor[HTML]{7D7DFF}{A}B and the parabola.

So first we recognize that if point \textcolor[HTML]{7D7DFF}{A}=$(-d,d)$,
then point B=$(1-d,1-d)$. Since line \textcolor[HTML]{7D7DFF}{A}B has an
equation in form $y=ax+b$, we can set up equations for points
\textcolor[HTML]{7D7DFF}{A} and B as follows: $d=a\cdot(-d)+b$ $(1)$ and
$1-d=a\cdot(1-d)+b$ $(2)$. Now $(1)-(2)$ results in $a=1-2d$ and thus, by using
$(1)$ again we get $b=2d-2d^2$.

Second, to obtain intersection point T we consider equation
$ax+b=\frac{x^2}{2}+\frac{1}{2}$ which can be reformulated to search the
roots of quadratic function $\frac{x^2}{2}-ax-b+\frac{1}{2}$. If and only
if the discriminant of this quadratic expression is zero, then
\textcolor[HTML]{7D7DFF}{A}B is a tangent. Indeed, the determinant is
$(-a)^2-4\cdot\frac{1}{2}\cdot(-b+\frac{1}{2})=a^2+2b-1$ which is, after
expanding $a$ and $b$, obviously zero.

Despite this is an analytical proof, by computing the $x$-coordinate of T
(which is $a=1-2d$) the reader may think of finding a synthetic proof as
well.

\subsection{Mathematics in the coffee cup}
\begin{figure}[H]
\begin{center}
\includegraphics[width=60mm]{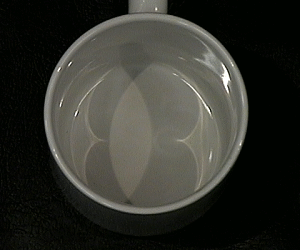}
\caption{The nephroid curve in the coffee cup.
A photo by Stuart Levy published at \url{http://www.geom.uiuc.edu/~fjw/calc-init/nephroid/}.}
\label{fig:15}
\end{center}
\end{figure}

The nephroid curve (see inside the coffee cup on
Fig.~\ref{fig:15}) is a 6th order algebraic curve defined by the envelope of a
set of mirrored light rays as a family of curves. Unfortunately,
computationally it is rather complex to solve the corresponding equation
system, thus it is inconvenient to use the {\bf Envelope} command with the recent
version of GeoGebra.

From the optical point of view, there are two approaches. One possibility
is to assume that the source of the light is a point. In this case the rays
are concurrent. The other possibility is to assume that the source is
infinitely distant, in this case the rays are parallel. Clearly, the
second case is the mathematical "limit" of the first one since if the point
converges to infinity, the models are closer and closer to each other.

\begin{figure}[H]
\begin{center}
\includegraphics[width=120mm]{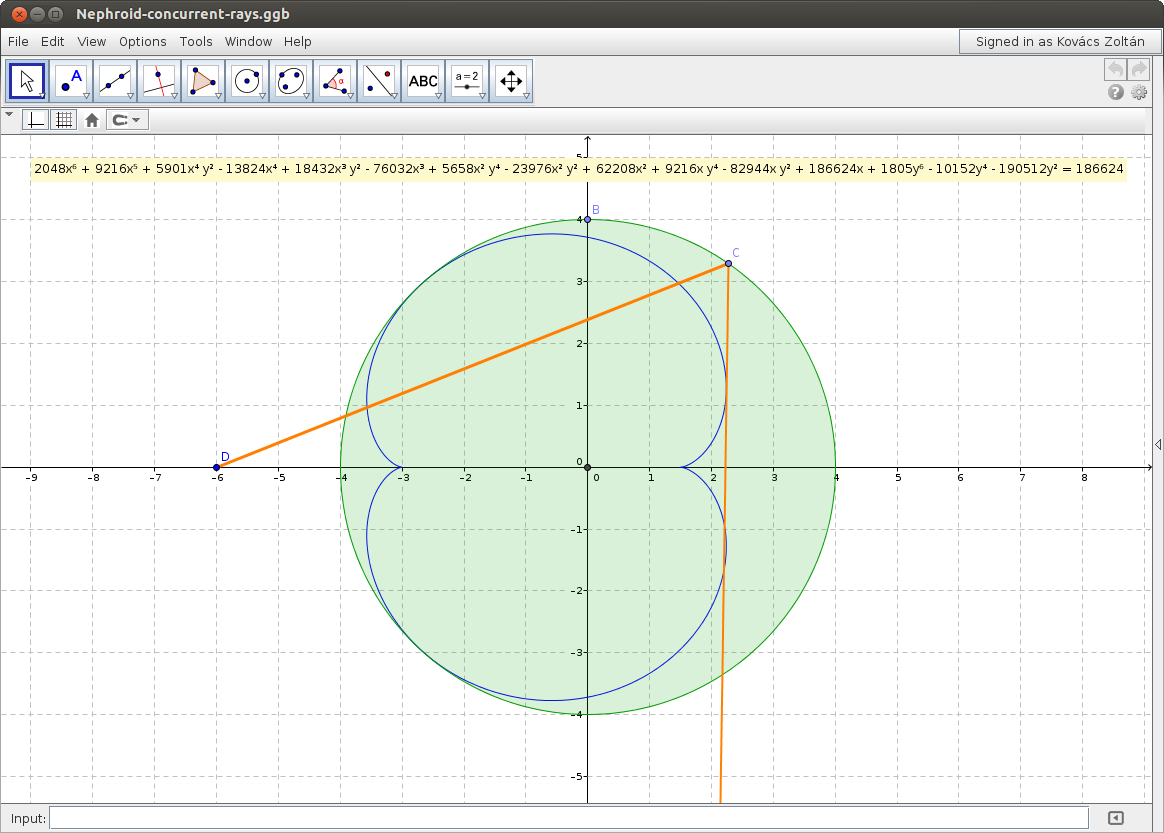}
\caption{A nephroid as computed by the desktop version of GeoGebra.}
\label{fig:16}
\end{center}
\end{figure}

The first approach is computationally easier. In Figure \ref{fig:16} we
can investigate the model of the concurrent rays by using the Java desktop
version of GeoGebra.

\begin{figure}[H]
\begin{center}
\includegraphics[width=120mm]{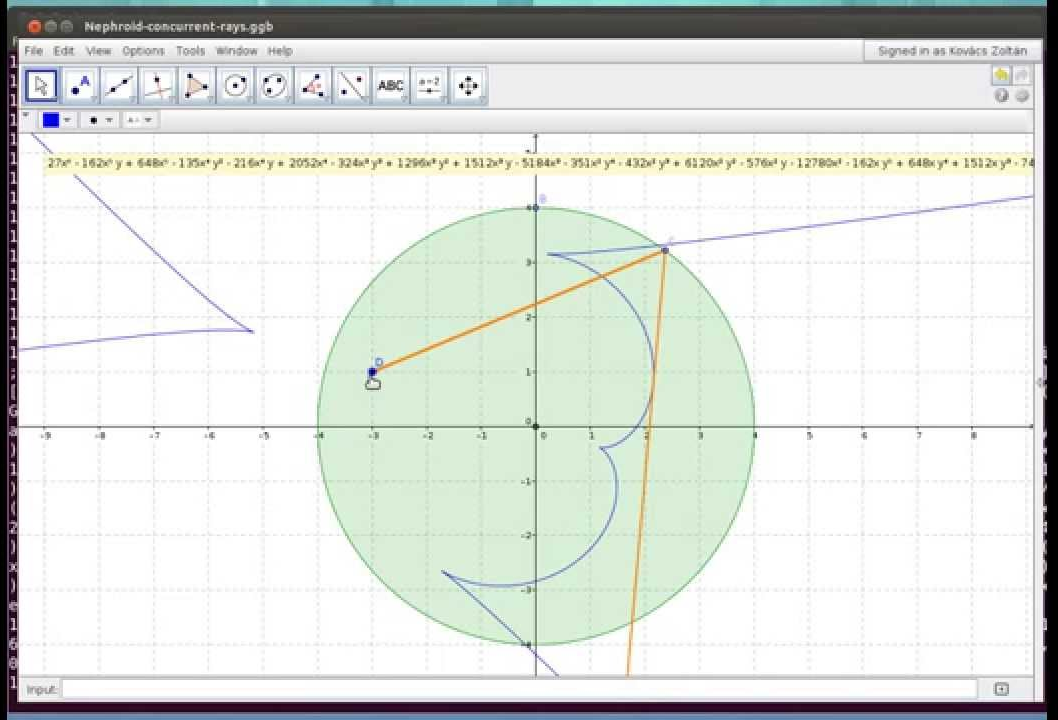}
\caption{A capture of the construction of a nephroid. YouTube video at \url{https://www.youtube.com/watch?v=nV_C4N7mWGs}.}
\label{fig:17}
\end{center}
\end{figure}

GeoGebra applets in GeoGebraBooks use Giac, but newest versions of GeoGebra
can be configured to use faster methods than Giac has. In a
video (Fig.~\ref{fig:17}) we can learn how the Java desktop version can be started to use or
not use the external computation machine SingularWS with the embedded
Gröbner cover algorithm.

Finally, the approach of the parallel rays is computationally the most
difficult one. It is impossible for Giac to compute the envelope equation
in a reasonable time, thus we have to force using SingularWS and the
Gröbner cover method by using the command line option {\tt {-}{-}singularws=enable:true}
again (Fig.~\ref{fig:18} shows a capture of the mentioned video):

\begin{figure}[H]
\begin{center}
\includegraphics[width=80mm]{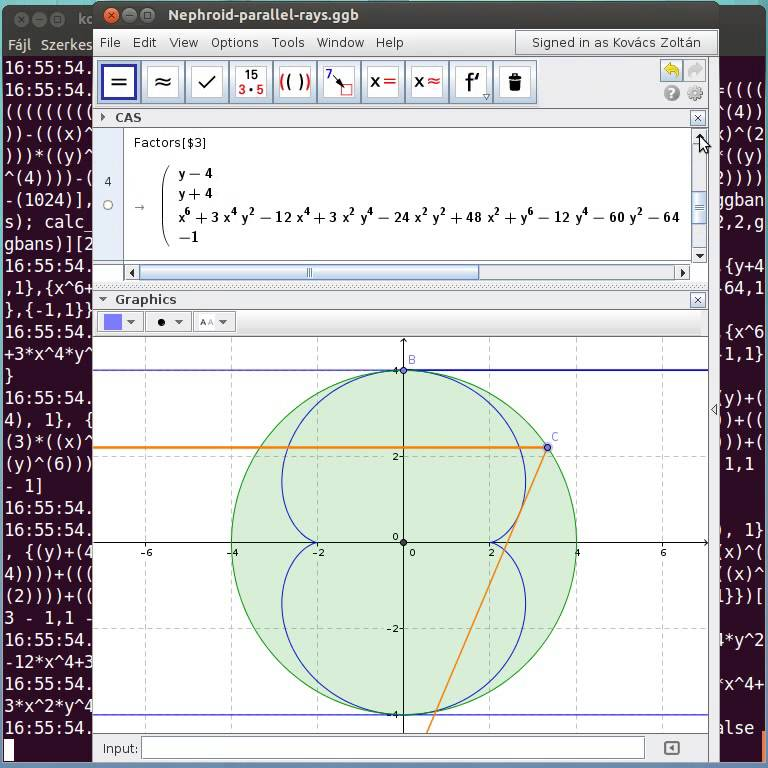}
\caption{The nephroid when rays are parallel. YouTube video at \url{https://www.youtube.com/watch?v=-mGTaJR2zyw}.}
\label{fig:18}
\end{center}
\end{figure}

As we can see in the above videos (Fig.~\ref{fig:17} and \ref{fig:18}), in the parallel case there are two extra
components which can be separated by factorization. But also in the
concurrent case when the source of the light is a perimeter point of the
circle there is an extra component.

\subsubsection*{Conclusion}

As \href{http://www.phikwadraat.nl/huygens_cusp_of_tea/}{Sander Wildeman} \textnormal{\cite{17}}
remarks, {\em once you have written an article about caustics you start to
see them everywhere}. We can only
agree: mathematics is everywhere, not only in geometric forms of basic
objects but various loci and envelopes. Mathematics is indeed everywhere---so
maths teachers can build motivation on emphasizing these not well known
facts in the modern era of education.

We are hoping that the newly developed GeoGebra commands {\bf LocusEquation} and {\bf Envelope}
will be a step forward in introducing and investigating real life objects in a
modern way for secondary school students. As future work, detailed study of these tools may be desirable in
concrete classroom situations.

\begin{figure}[H]
\begin{center}
\includegraphics[width=80mm]{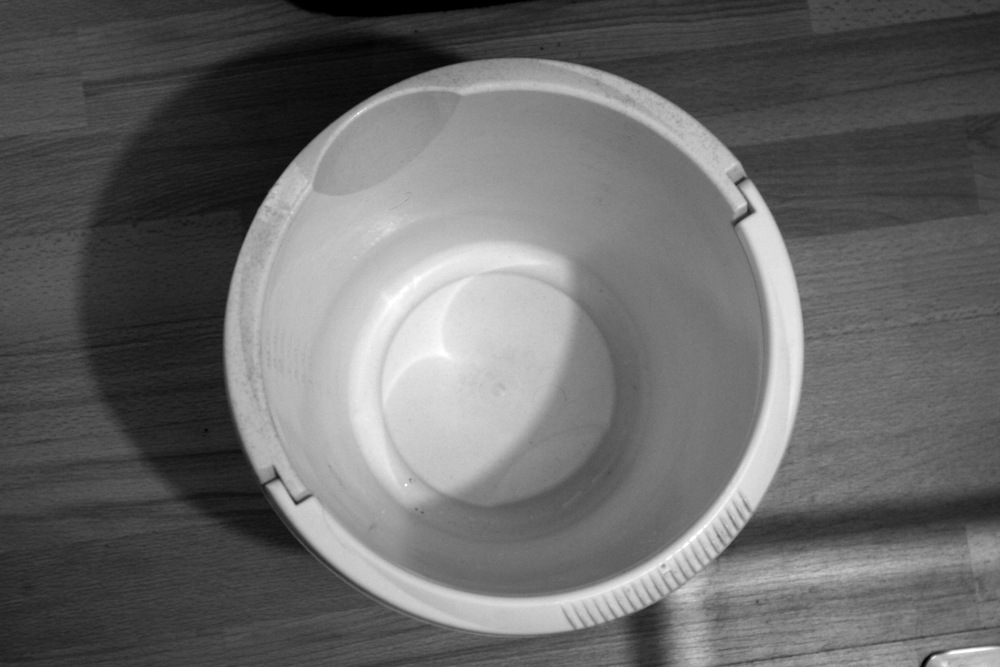}
\caption{Mathematics is everywhere.}
\label{fig:19}
\end{center}
\end{figure}

\section{Acknowledgments}
First author partially supported by the Spanish ‘‘Ministerio de
Economía y Competitividad’’ and by the European Regional
Development Fund (ERDF), under the Project MTM2011-25816-C02-02.

\end{document}